# ReconU-Net: a direct PET image reconstruction using U-Net architecture with back projection-induced skip connection

Fumio Hashimoto[1] and Kibo Ote[1]

[1] Central Research Laboratory, Hamamatsu Photonics K. K., 5000 Hirakuchi, Hamakita-ku, Hamamatsu 434-8601, Japan
E-mail: fumio.hashimoto@crl.hpk.co.jp

**Abstract**

[Objective] This study aims to introduce a novel back projection-induced U-Net-shaped architecture, called ReconU-Net, for deep learning-based direct positron emission tomography (PET) image reconstruction. Additionally, our objective is to analyze the behavior of direct PET image reconstruction and gain deeper insights by comparing the proposed ReconU-Net architecture with other encoder-decoder architectures without skip connections.
[Approach] The proposed ReconU-Net architecture uniquely integrates the physical model of the back projection operation into the skip connection. This distinctive feature facilitates the effective transfer of intrinsic spatial information from the input sinogram to the reconstructed image via an embedded physical model. The proposed ReconU-Net was trained using Monte Carlo simulation data from the Brainweb phantom and tested on both simulated and real Hoffman brain phantom data.
[Main results] The proposed ReconU-Net method generated a reconstructed image with a more accurate structure compared to other deep learning-based direct reconstruction methods. Further analysis showed that the proposed ReconU-Net architecture has the ability to transfer features of multiple resolutions, especially non-abstract high-resolution information, through skip connections. Despite limited training on simulated data, the proposed ReconU-Net successfully reconstructed the real Hoffman brain phantom, unlike other deep learning-based direct reconstruction methods, which failed to produce a reconstructed image.
[Significance] The proposed ReconU-Net can improve the fidelity of direct PET image reconstruction, even when dealing with small training datasets, by leveraging the synergistic relationship between data-driven modeling and the physics model of the imaging process.

## 1. Introduction

Positron emission tomography (PET) is a molecular imaging technique utilized for visualizing and quantifying the distribution of PET tracers in living humans [1]. Due to its versatility, PET has been used not only for cancer detection [2] and neurodegenerative disease diagnosis, such as Alzheimer's disease [3], but also in fundamental research [4]. While PET stands out as a unique imaging modality capable of tracking picomole-order molecules, image noise is more pronounced compared to other tomography scanners, such as X-ray computed tomography (CT), owing to the limited statistical counts in the acquired data. The presence of image noise compromises quantitative accuracy and lesion detectability, potentially leading to the unfortunate scenario of missed lesions. Therefore, noise reduction techniques are essential for low-dose or short-time scans.

To reduce statistical noise in PET images, iterative reconstruction algorithms using various regularizations have been developed. The classical approach to penalized PET image reconstruction involves measuring spatial smoothness in the reconstructed image space using Gibbs priors [5-7]. With the development of PET/CT and PET/magnetic resonance imaging (MRI) scanners, several penalized PET image reconstruction algorithms incorporating additional anatomical information from CT or MR images have also been developed [8-10]. More recently, the emergence of deep learning has brought about a paradigm shift in PET image reconstruction [11-13].

Incorporating deep learning as a penalty for iterative image reconstruction has been reported to enhance image quality and has the potential to push the limitations of existing iterative image reconstruction algorithms [14-18]. Alternatively, a direct PET image reconstruction approach has also been proposed, in which the relationship between measurement data and reconstructed images is obtained in a data-driven manner, attracting attention due to its high calculation speed [19-24]. This is calculated with only a single forward path, differing from iterative reconstruction, which repeats the forward and back-projection processes. The first attempt at direct medical image reconstruction was the automated transform by manifold approximation (AUTOMAP) by Zhu et al., which introduced dense connections in the first and second layers of a neural network structure to acquire direct mapping from sinograms to reconstructed images [22]. Inspired by the AUTOMAP methods, Häggström et al. proposed the DeepPET method using a fully convolutional neural network (FCN) [23]. DeepPET consists of an encoder-decoder structure with improvements to address the challenges of utilizing FCNs for direct image reconstruction, such as larger convolution filter kernel sizes and a deeper layered network structure. Furthermore, direct PET image reconstruction methods have undergone several modifications in the loss function [24] and extensions to the long-axial field-of-view PET scanners [25].

While these direct PET image reconstruction methods yield visually appealing images reconstructed from sinograms, accurately obtaining the inverse transformation from sinograms to reconstructed images through a data-driven approach remains challenging. These direct reconstruction methods may produce "false-structured" finer PET images.

In this study, we propose a novel back projection-induced U-Net-shaped architecture, called ReconU-Net, for direct PET image reconstruction. The proposed ReconU-Net architecture distinctly integrates the physical model of the back-projection operation into the skip connection, facilitating the effective transfer of intrinsic spatial information from the input sinogram to the reconstructed image through the embedded physical model. This innovative architecture aims to enhance the fidelity of direct PET image reconstruction by capitalizing on the synergistic relationship between data-driven modeling and the physics of the imaging process. Additionally, we offer further insights by comparing the proposed ReconU-Net architecture with other encoder-decoder architectures without skip connections.

## 2. Methodology

### 2.1 Direct PET image reconstruction

Direct image reconstruction methods, such as DeepPET [23], employ an encoder-decoder type CNN with a deeper layer network structure to obtain the reconstructed image *x* from sinogram *y* through latent features in the bottleneck layer. This process is expressed as:

$$\theta^* = \underset{\theta}{\mathrm{argmin}} \frac{1}{N} \sum_{i \in D} E(f(\theta_{Enc}, \theta_{Dec}|y_i); x_i), \tag{1}$$

where *f* is the encoder-decoder network with trainable parameters $\theta_{Enc}$ and $\theta_{Dec}$ of the encoder and decoder, respectively; *E* is the loss function, such as the mean squared error (MSE), and *D* is a mini-batch sample of size *N*. In this direct reconstruction method, the intrinsic spatial information in sinograms faces a challenge in seamless transfer from the encoder to the decoder through skip connections. This is because the encoder and decoder operate in different spaces, namely, the sinogram and reconstructed image spaces, respectively. This inherent limitation in the network structure hampers the performance of direct image reconstruction.

### 2.2 Proposed method

In this study, we introduced the ReconU-Net architecture, designed to facilitate the smooth transfer of spatial information from the input sinogram in the encoder to the decoder. An overview of the proposed ReconU-Net architecture is depicted in Figure 1. ReconU-Net guides the connections of features from the encoder, expressed in sinogram space, to the decoder, expressed in the reconstructed image space, using each back-projection module in the skip connections. The *k*-th scale feature map after the back-projection operation $\hat{\Gamma}$ is represented as follows.

$$\hat{\Gamma}_k = \sum_{i=1}^{I_k} a_{k_{ij}} \Gamma^c_{k,i} \tag{2}$$

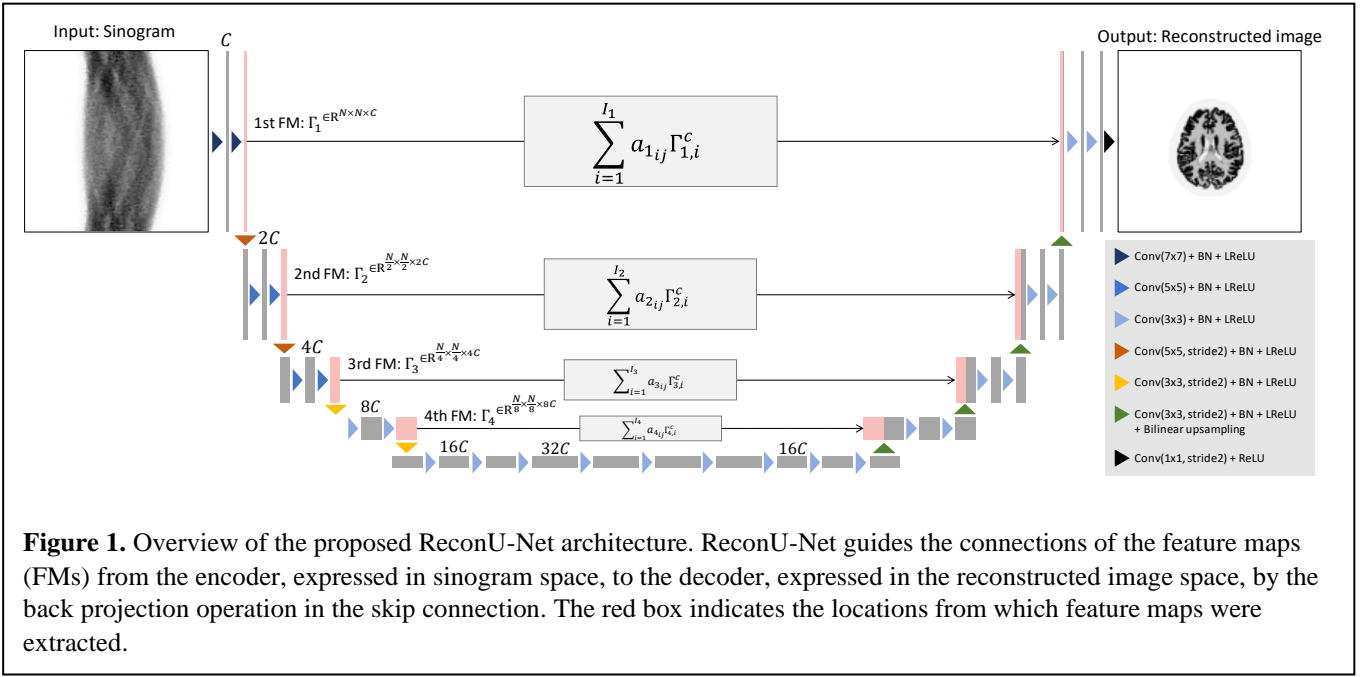

**Figure 1.** Overview of the proposed ReconU-Net architecture. ReconU-Net guides the connections of the feature maps (FMs) from the encoder, expressed in sinogram space, to the decoder, expressed in the reconstructed image space, by the back projection operation in the skip connection. The red box indicates the locations from which feature maps were extracted.

where $\Gamma_k$ and $a_k$ are the k-th scale feature map from the encoder part and system matrix, respectively. *where i, j*, and *c* are the indices of the line-of-response, voxel, and feature maps, respectively. We concatenate $\hat{\Gamma}$ to each scale decoder part in this architecture. Thus, the proposed ReconU-Net architecture retains accurate spatial information from the input sinograms, aligning with the physical model.

The network architecture of the proposed method mirrors that of the DeepPET network, differing mainly in the skip connections. In the encoder, a 2D convolution layer with batch normalization (BN) and a leaky rectified linear unit (LReLU) was repeated twice, followed by a 2D convolution layer with two strides for downsampling, succeeded by the BN and LReLU. The convolution filter kernels for each scale were reduced to 7×7, 5×5, and 3×3, as depicted in Figure 1. The number of feature maps was doubled at each downsampling step, with $N$ = 128 and C = 32, indicating the input sinogram size and the number of feature maps, respectively. At the bottleneck of the encoder, 1024 feature maps of $8 \times 8$ feature maps were obtained. In the decoder, a 2D convolution layer with BN and LReLU was employed, followed by a bilinear upsampling and concatenation operation from the skip connection. Subsequently, the 2D convolution layer with BN and LReLU was repeated twice, with the number of feature maps halved after each concatenation operation.

For network training, MSE was utilized as the loss function, and Adam, with a learning rate of 1e-03, served as the optimizer. The number of epochs and batch sizes were set to 500 and 70, respectively. The training was executed on a workstation running Ubuntu 20.04, equipped with a graphics processing unit of NVIDIA A100 with 80 GB of memory, and PyTorch 1.12.1.

## 3. Experimental setup

### 3.1 Simulation data generation

We generated twenty sets of 3D digital brain PET data from the BrainWeb phantom (https://brainweb.bic.mni.mcgill.ca/brainweb/) using Monte Carlo simulation, incorporating the specific geometry of a brain-dedicated PET scanner [26]. The radioactivity contrast between gray matter, white matter, and cerebrospinal fluid was set at a ratio of 1:0.25:0.05, reflecting the [$^{18}$F]FDG distribution. For each subject, we conducted a 3D data acquisition, resulting in a total of 181.12 ± 6.08 million counts, inclusive of scatter events.

The simulation data were divided into 18 subjects for training, one subject for validation, and one subject for testing. 2D sinograms were generated from the list-mode data using a single-slice rebinning (SSRB) method with a maximum ring difference of 15. The phantom and sinogram sizes were 128 × 128 voxels with a voxel size of 3.0 × 3.0 mm$^2$, and 128 bins × 128 angles, respectively. Corrections were applied in the sinogram space prior to image reconstruction.

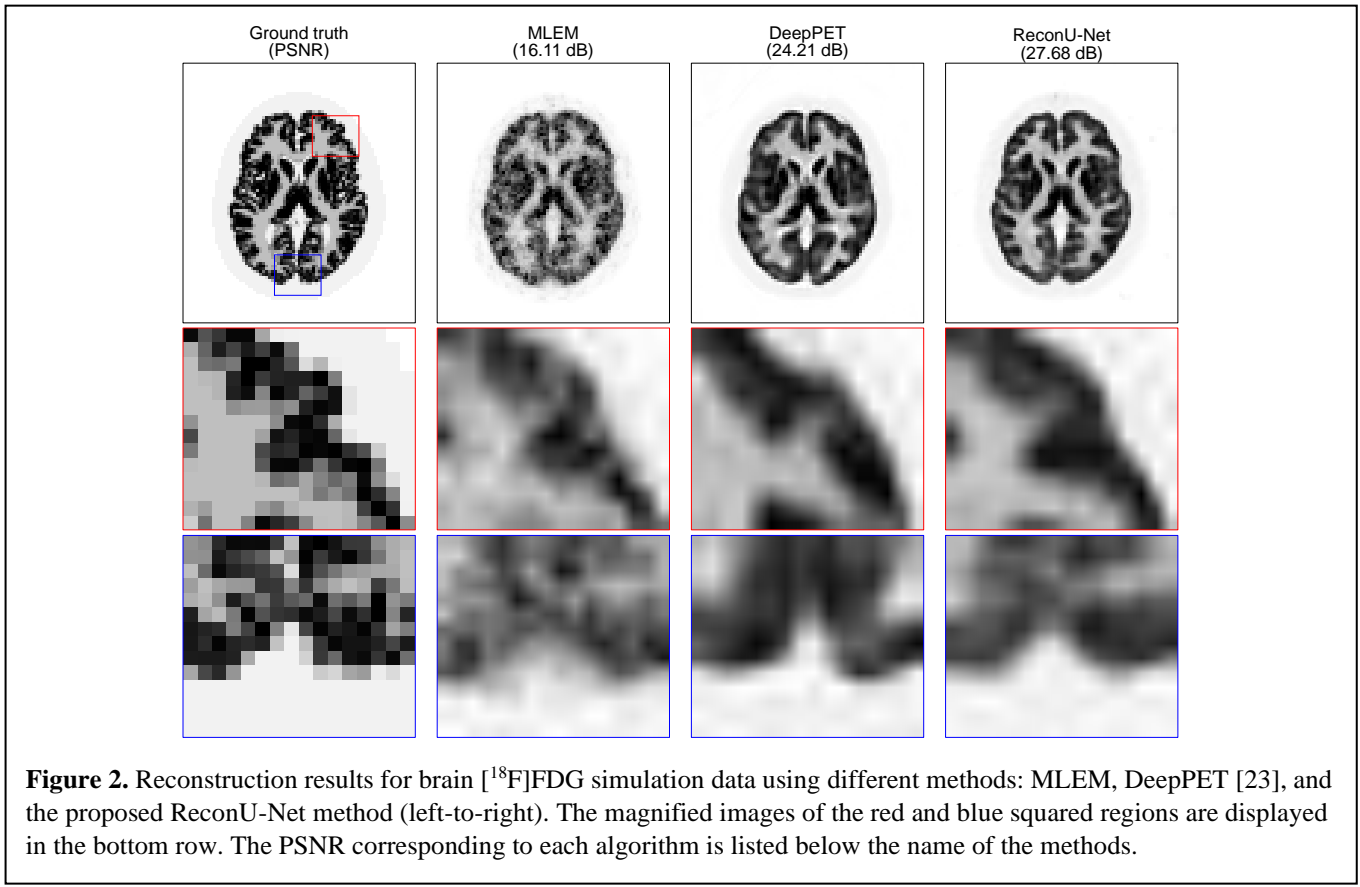

**Figure 2.** Reconstruction results for brain [$^{18}$F]FDG simulation data using different methods: MLEM, DeepPET [23], and the proposed ReconU-Net method (left-to-right). The magnified images of the red and blue squared regions are displayed in the bottom row. The PSNR corresponding to each algorithm is listed below the name of the methods.

*3.2 Application of real phantom data*

The proposed model, trained using simulation data, was applied to reconstruct the Hoffman brain phantom obtained from a brain-dedicated PET scanner [23]. A 2,000-second emission scan was conducted with 21.3 MBq of 18-F. 2D sinograms were generated from list-mode data using the SSRB method with a maximum ring difference of 15, and corrections were applied in the sinogram space before image reconstruction. The sizes of the reconstructed image and sinogram were 128 × 128 voxels with a voxel size of 3.0 × 3.0 mm$^2$, and 128 bins × 128 angles, respectively. Additionally, a 3D list-mode image reconstruction with a maximum ring difference of 66 and time-of-flight of 300 ps was performed using the dynamic row action maximum likelihood algorithm (DRAMA) [27] for the reference image.

*3.3 Evaluation*

We compared the performance of the proposed ReconU-Net with that of the maximum likelihood expectation maximization (MLEM) and DeepPET [23] methods. The numbers of trainable parameters for the proposed ReconU-Net and DeepPET architectures are approximately 55 million and 54 million, respectively. Note that the DeepPET architecture is a version of the proposed ReconU-Net architecture without skip connections.

The peak signal-to-noise ratio (PSNR) was calculated to evaluate image quality as follows:

$$PSNR = 10\, log_{10}\left(\frac{max(K)^2}{\frac{1}{N}\|K - K'\|_2^2}\right), \tag{3}$$

where $K$ and $K'$ are the ground truth and target reconstructed image, $N$ is the number of voxels, and $max\,(\cdot)$ is the maximum value of the image.

**4. Results and Discussion**

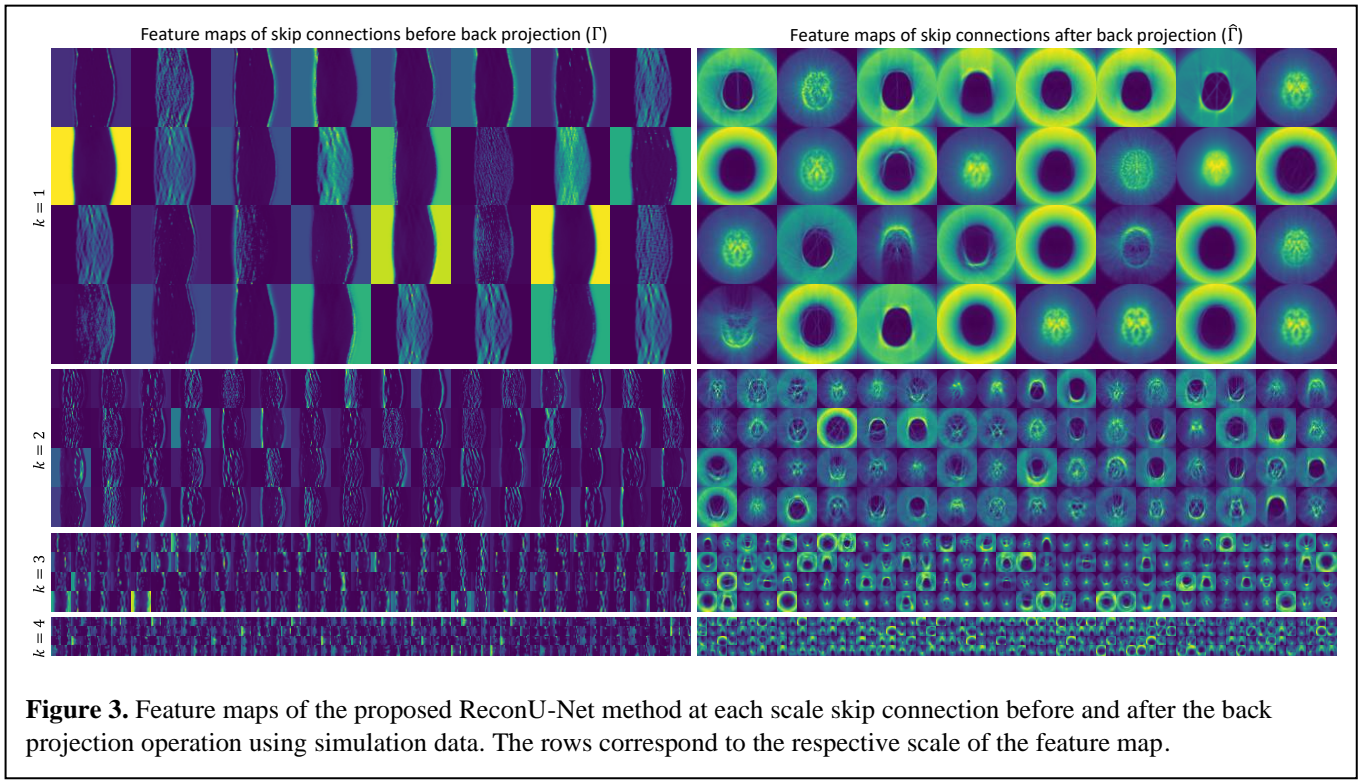

**Figure 3.** Feature maps of the proposed ReconU-Net method at each scale skip connection before and after the back projection operation using simulation data. The rows correspond to the respective scale of the feature map.

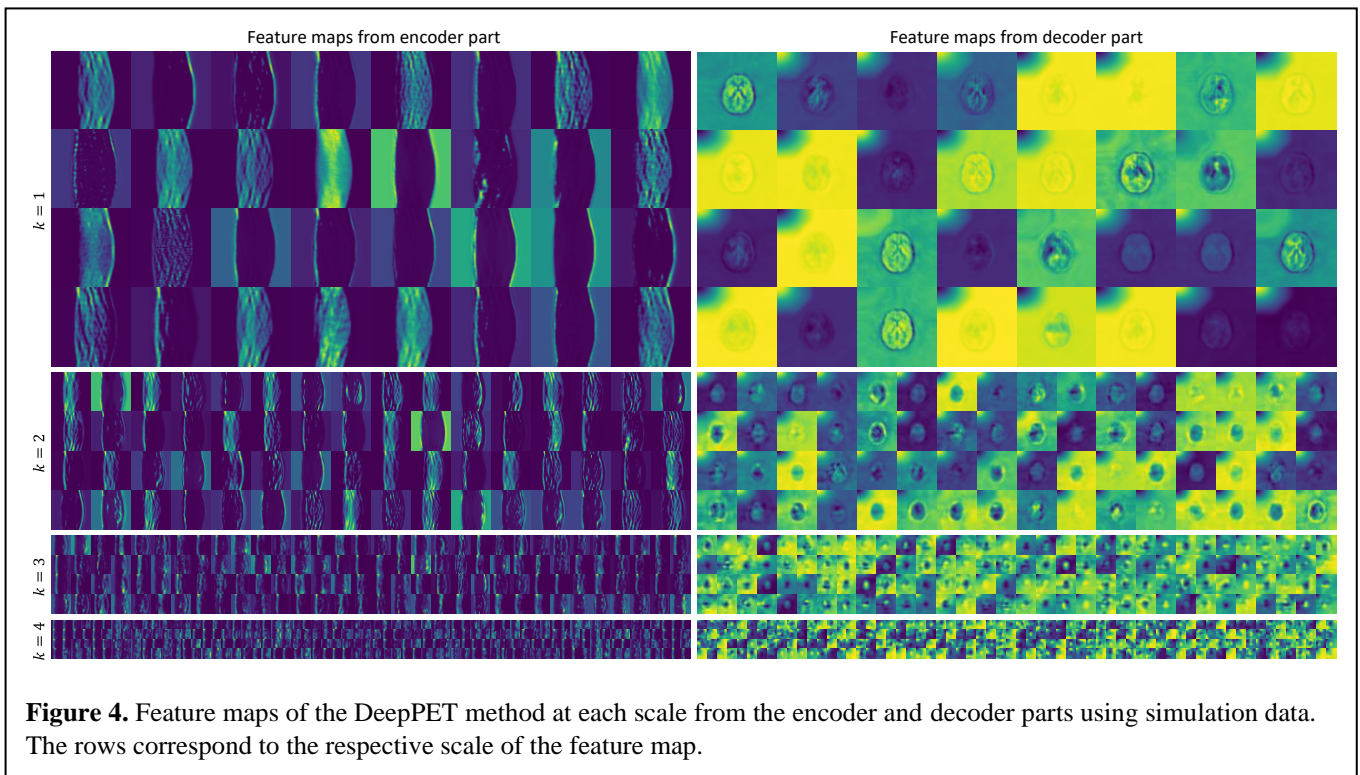

**Figure 4.** Feature maps of the DeepPET method at each scale from the encoder and decoder parts using simulation data. The rows correspond to the respective scale of the feature map.

Fig. 2 shows the reconstruction results of the brain [$^{18}$F]FDG simulation data for different methods. The reconstructed image of the MLEM was performed with the Gaussian post-filter of σ=1 voxel. Among these methods, the proposed ReconU-Net yielded the highest PSNR in the simulation. In comparison with the DeepPET method, the proposed ReconU-Net method generated a reconstructed image with a more accurate structure. This improvement is attributed to the capability of the proposed ReconU-Net architecture to seamlessly transfer intrinsic spatial information from the input sinogram, achieved by explicitly incorporating the back-projection operation into the network structure.

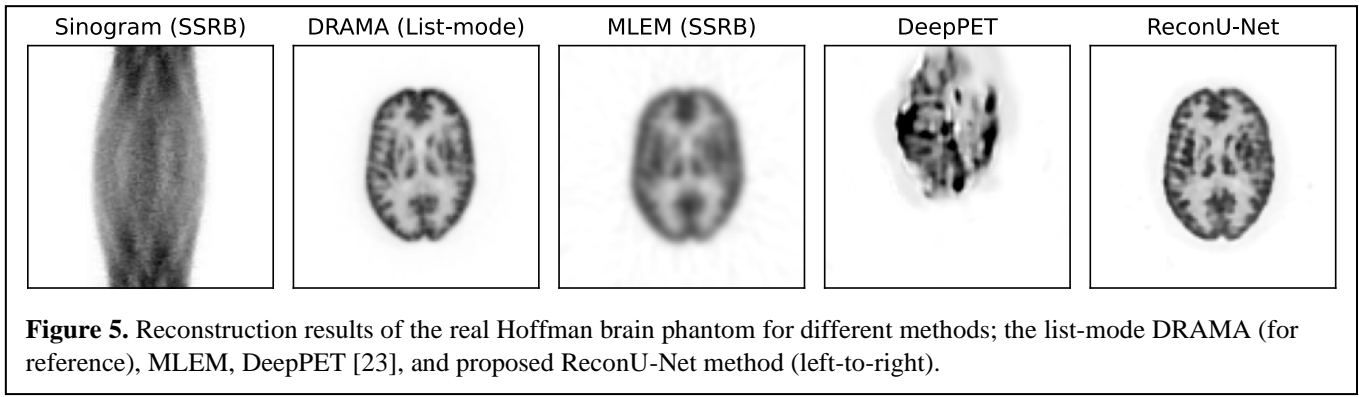

**Figure 5.** Reconstruction results of the real Hoffman brain phantom for different methods; the list-mode DRAMA (for reference), MLEM, DeepPET [23], and proposed ReconU-Net method (left-to-right).

Fig. 3 displays the feature maps of the proposed ReconU-Net method at each scale skip connection before and after the back projection operation $\Gamma$ and $\hat{\Gamma}$ using simulation data. The first-scale ($k$=1) feature maps generated head-shaped images, sinograms filtered by reconstruction-like filters, and their corresponding back-projected images. The second-scale ($k$=2) feature map produced images with geometric patterns and segmented white matter. These findings suggest that the proposed ReconU-Net architecture efficiently learns more straightforward reconstruction filtering in shallow layers, such as first-scale features, and more abstract features in deeper layers, such as second- and subsequent-scale features. Fig. 4 displays the feature maps of the DeepPET method at each scale from the encoder and decoder parts using simulation data. The DeepPET architecture employed in this study and the locations from which the feature maps were extracted are shown in Supplementary Figure 1. The encoder component of DeepPET exhibits characteristics similar to those of ReconU-Net. However, these architectures produced widely different feature maps of the decoder, which explicitly transferred intrinsic spatial information from the encoder to the decoder by guiding the physical model using a back-projection operation. However, the DeepPET architecture produced blurred feature maps as it had to reconstruct from latent features in the bottleneck layer. In contrast, the proposed ReconU-Net architecture can efficiently transfer multiple-resolution features, especially nonabstract high-resolution information, through skip connections. It is also anticipated to mitigate the vanishing-gradient problem and expedite the learning process, akin to the original U-Net architecture [28, 29].

Fig. 5 shows the reconstruction results for a real Hoffman brain phantom. In the real data experiment, a phantom image reconstructed using the 3D list-mode DRAMA algorithm without any blurring from the SSRB method served as the reference phantom image. Despite being trained only on a limited amount of simulated data, the proposed method successfully generated a reconstructed image. In contrast, the DeepPET network, also trained on a restricted set of simulated data, failed to produce a reconstructed image from the real PET data. This is primarily attributed to the fact that DeepPET requires an extensive training dataset for model generalization. While our proposed ReconU-Net utilized a 1,260 sinogram-phantom image training dataset, the original DeepPET employed approximately 160 times more, totaling 203,305 sinogram-phantom image training datasets. The results demonstrate that the proposed ReconU-Net architecture can enhance the fidelity of direct PET image reconstruction, even with small datasets, by capitalizing on the synergistic relationship between data-driven modeling and the physics of the imaging process. In other words, generating images from deep latent features has many pitfalls, but multiresolution image reconstruction from back-projected feature maps induced by skip connections may be easier to obtain.

The principal limitation of this study was that our evaluation was based solely on the brain [$^{18}$F]FDG simulation dataset and real Hoffman phantom data using 2D PET sinograms. Future studies will necessitate more comprehensive evaluations using an expanded set of training and testing data. It is also essential to consider detailed quantitative analyses, such as region-of-interest analyses. Additionally, we explored the performance of the proposed ReconU-Net in low-dose PET imaging.

## 5. Conclusion

In this study, we introduced a novel back projection-induced ReconU-Net architecture for direct PET image reconstruction. The proposed ReconU-Net architecture uniquely integrates the back-projection operation into the skip connection, facilitating the transfer of intrinsic spatial information from the input sinogram to the reconstructed PET image through the physical model. The experiments demonstrated that the proposed ReconU-Net method generated a reconstructed image with a more accurate structure than the DeepPET method. Furthermore, more in-depth analyses showed that the proposed ReconU-Net could enhance the fidelity of direct PET image reconstruction, even when dealing with small training datasets, by leveraging the synergistic relationship between data-driven modeling and the physical model of the imaging process.

**Acknowledgments**

The authors would like to thank the expert assistance and advice provided by Mr. Etsuji Yoshikawa, Mr. Takashi Isobe, and Mr. Yuya Onishi from the Central Research Laboratory, Hamamatsu Photonics K.K.

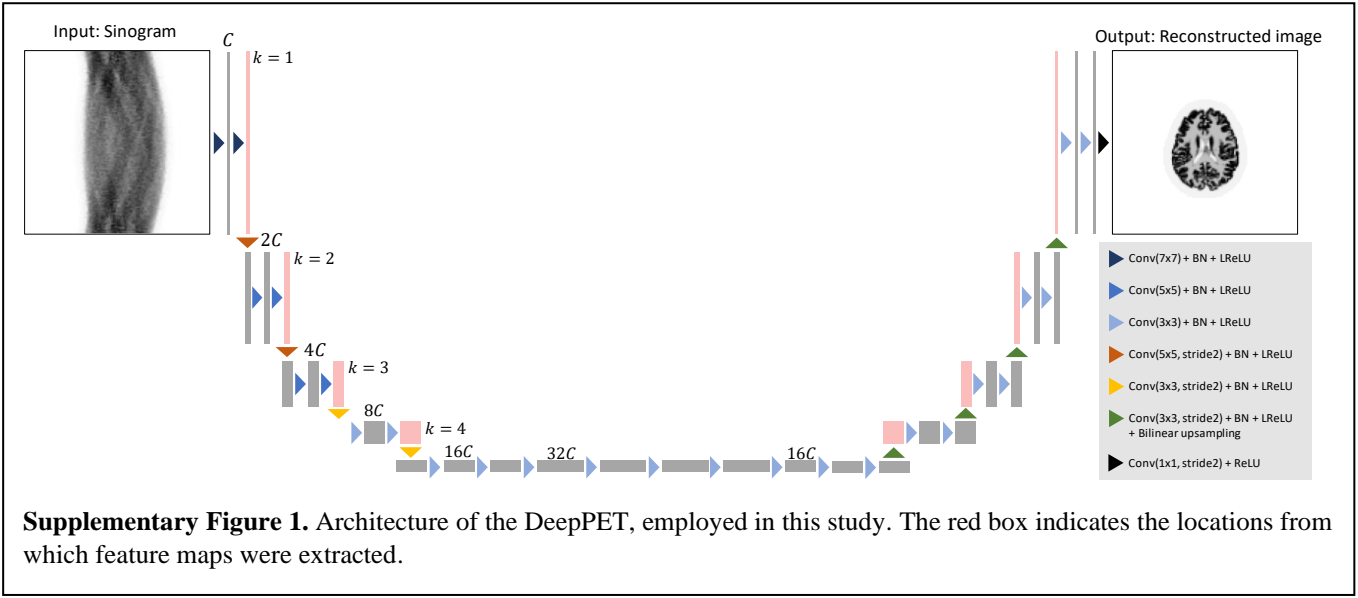

**Supplementary Figure 1.** Architecture of the DeepPET, employed in this study. The red box indicates the locations from which feature maps were extracted.